\documentclass[twocolumn]{aastex62}
\usepackage{graphicx,epsfig,amssymb,amsmath,color,lineno}
\usepackage{makecell}

\newcommand{\redcheck}{{\color{black}\checkmark}}
\newcommand{\myhash}{\raisebox{\depth}{\#}}

\shortauthors{Pe\~nil et al.}

\begin{document}

\title {Systematic search for $\gamma$-ray periodicity in active galactic nuclei detected by the {\it Fermi} Large Area Telescope}

\email{E-mails: ppenil@ucm.es, alberto.d@ucm.es}

\author{P. Pe\~nil}
\affil{IPARCOS and Department of EMFTEL, Universidad Complutense de Madrid, E-28040 Madrid, Spain}

\author{A. Dom\'inguez}
\affil{IPARCOS and Department of EMFTEL, Universidad Complutense de Madrid, E-28040 Madrid, Spain}

\author{S. Buson}
\affil{Julius-Maximilians-Universit\"at, 97070, W\"urzburg, Germany}

\author{M. Ajello}
\affil{Department of Physics and Astronomy, Clemson University,	Kinard Lab of Physics, Clemson, SC 29634-0978, USA}

\author{J. Otero-Santos}
\affil{Instituto de Astrofísica de Canarias, E-38200 La Laguna, Tenerife, Spain}

\author{J.~A. Barrio}
\affil{IPARCOS and Department of EMFTEL, Universidad Complutense de Madrid, E-28040 Madrid, Spain}

\author{R. Nemmen}
\affil{Universidade de São Paulo, Instituto de Astronomia, Geofísica e Ciências Atmosféricas,\\ Departamento de Astronomia SP 05508-090 São Paulo, Brazil}

\author{S. Cutini}
\affil{Istituto Nazionale di Fisica Nucleare, Sezione di Perugia, I-06123 Perugia, Italy}

\author{B. Rani}

\affil{Southeastern Universities Research Association, Washington DC, USA}
\affil{NASA Goddard Space Flight Center, Greenbelt, MD 20771, USA}
\affil{Center for Research and Exploration in Space Sciences and Technology, NASA/GSFC, Greenbelt, MD 20771, USA}
\affil{Korea Astronomy \& Space Science Institute, 776, Daedeokdae-ro, Yuseong-gu, Daejeon, 305-348, Republic of Korea}

\author{A. Franckowiak}
\affil{Deutsches Elektronen-Synchrotron DESY Platanenallee 6, D-15738, Zeuthen, Germany}

\author{E. Cavazzuti}
\affil{Italian Space Agency, Via del Politecnico snc, 00133 Roma, Italy}



\begin{abstract}

We use nine years of $\gamma$-ray data provided by the {\it Fermi} Large Area Telescope (LAT) to systematically study the light curves of more than two thousand active galactic nuclei (AGN) included in recent {\it Fermi}-LAT catalogs. Ten different techniques are used, which are organized in an automatic periodicity-search pipeline, in order to search for evidence of periodic emission in $\gamma$ rays. Understanding the processes behind this puzzling phenomenon will provide a better view about the astrophysical nature of these extragalactic sources. However, the observation of temporal patterns in $\gamma$-ray light curves of AGN is still challenging. Despite the fact that there have been efforts on characterizing the temporal emission of some individual sources, a systematic search for periodicities by means of a full likelihood analysis applied to large samples of sources was missing. Our analysis finds 11 AGN, of which 9 are identified for the first time, showing periodicity at more than $4\sigma$ in at least four algorithms. These findings will help in solving questions related to the astrophysical origin of this periodic behavior.
\end{abstract}



\section{Introduction} \label{sec:intro}
One conclusion after decades of multiwavelength observations is that supermassive black holes (SMBHs) are present at the centers of most galaxies \citep[e.g.,~][]{soltan_1982, cavaliere_1989, chokshi_1992}. A fraction of these galactic centers has the environmental conditions in terms of a dense accretion disk to feed the SMBH, transforming such objects into powerful emitters known as AGN \citep[e.g.,~][]{wiita_lecture}. These emissions are characterized by variability on different time scales and can emerge in the form of relativistic jets in some sources \citep[e.g.,~][]{sartori_variability}. 
The light curves (LCs) of these sources show temporal behavior, which may or may not display a specific pattern. For instance, finding periodic emission in a source can provide information about its astrophysical nature. Possible explanations of periodic behavior in AGN emissions are lighthouse effects in jets \citep[e.g.~][]{camenzind_jet}, modulations in the accretion flow \citep[e.g.~][]{gracia_accretion}, or the existence of binary SMBHs \citep[e.g.,~][]{sobacchi_binary, celoria_binary}. Furthermore, predictions of future flux modulations may help in scheduling more efficient observations with Imaging Atmospheric Cerenkov Telescopes, which have limited fields of view and observing duty cycles \citep[e.g.,~VERITAS, MAGIC, H.E.S.S.;][respectively]{weekes_veritas,magic_lorenz,hinton_hess}.

Different strategies have been employed in the literature to detect periodicities in the $\gamma$-ray LCs of AGN. The typical approach consists of analyzing one object by applying a few time series algorithms, with a minimum of two algorithms for cross-checking results. There are studies where this cross check is complemented with cross-correlation of data at other wavelengths \citep[e.g.,~][]{ackermann_pg1553}. However, \cite{prokhorov_set} and \cite{covino_negation} analyzed a sample of 7 and 10 AGN, respectively, to search for gamma-ray periodicity. The former work determined their sample through the search for periodicities in the entire sky and found a periodicity in 7 objects, whereas the latter found no evidence of periodicity for any of their studied sources.


In general, previous works were based on individual studies. Employing data taken by the \textit{Fermi} Large Area Telescope (LAT) during the last decade and thanks to its all-sky scanning mode operation with complete sky coverage several times per day, we perform a systematic search for detecting periodic $\gamma$-ray emission in a sample containing $\sim$ 2000 AGN.

The paper is organized as follows. In $\S$\ref{sec:fermidata}, the AGN sample is presented, with an explanation of the process for analyzing the data provided by \textit{Fermi}-LAT. Then, $\S$\ref{sec:methodology} details the periodicity analysis methodology, first introducing the algorithms and techniques used for the periodicity detection and second describing the periodicity-search pipeline. We discuss in $\S$\ref{sec:candidates} the results obtained in our study, followed by a summary in $\S$\ref{sec:conclusion}.

\section{Gamma-ray sample} \label{sec:fermidata}
In this section we describe the AGN sample used in this work. We also present the procedure for generating the AGN LCs from the data provided by {\it Fermi}-LAT.

\subsection{Source Selection}
The \textit{Fermi} Gamma-ray Space Telescope was launched in June 2008 \citep{fermi_lat}. Its main instrument, LAT, is a pair-conversion detector that measures high-energy $\gamma$ rays with energies ranging from about 20 MeV to more than 300 GeV. LAT features a large field of view ($> 2$~sr) that allows scanning the entire sky in hours and therefore monitoring thousands of objects unbiased for spatial selections. LAT's all-sky monitoring capabilities provide us with long-coverage observations at different timescales, from seconds to years. Since 2008 almost continuous observations are available for a large number of $\gamma$-ray sources.

In this work, we utilize 28-days binned $\gamma$-ray light curves, computed at energies above 1~GeV, for more than 2000 {\it Fermi}-LAT AGN. We adopted a one month time bin, which is typically a good compromise between a computationally manageable program and sensitivity to long term variations.

The source sample is based on the combination of three {\it Fermi}-LAT catalogs: 3FGL \citep{3fgl_catalog}, 2FHL \citep{2fhl_catalog} and 3FHL \citep{3fhl_catalog}. 3FGL contains 3030 sources characterized in the 100 MeV--300 GeV range, based on the initial 4 years of the LAT activity. Regarding extragalactic sources, blazars (AGN with their jets aligned towards our line of sight) are the most numerous class, containing more than 1100 sources. 2FHL includes 360 objects detected above 50 GeV and characterized up to 2 TeV in the first 6.7 years of exposure. About 75\% of the sources in the catalog (274 sources) are extragalactic, and indeed the great majority are blazars. The 3FHL catalog reports sources detected at energies above 10 GeV, using the first 7 years of \textit{Fermi}-LAT data. 3FHL contains 1556 sources characterized up to 2 TeV. Most of the 3FHL sources ($\geq$ 79\%) are associated with extragalactic counterparts and in particular blazars. Combining the AGN in these catalogs, we obtain an initial sample of 2274 AGN. 

\subsection{{\it Fermi}-LAT Data Analysis}
The data analysis was performed following the {\it Fermi}-LAT collaboration recommendations for point-source analysis\footnote{\url{http://fermi.gsfc.nasa.gov/ssc/data/analysis/documentation/Pass8_usage.html}}, and is briefly outlined in the following. LAT data of the Pass~8 source class were selected spanning the time interval from August 2008 to October 2017 and analysed using the  {\it Fermi}-LAT ScienceTools package version v11r05p3 available from the \textit{Fermi} Science Support Center\footnote{\url{http://fermi.gsfc.nasa.gov/ssc/data/analysis/}} (FSSC) and the \textsf{P8R2\_SOURCE\_V6} instrument response functions, along with the fermipy software package \citep{Wood:2017yyb}. To minimize the contamination from $\gamma$-rays produced in the Earth's upper atmosphere, a zenith angle cut of $\theta<90\degr$ was applied. We applied also the standard data quality cuts (\textit{$DATA\_QUAL>0$}) and (\textit{$LAT\_CONFIG==1$}) and removed time periods coinciding with solar flares and $\gamma$-ray bursts detected by the LAT. For each source, we selected a $10\degr \times 10\degr$ region of interest centered at its catalog position (using RA and Dec) and photons with energies $>$1~GeV. The low-energy threshold of 1~GeV was driven mainly by computational limitation. The $\gamma$-ray flux in each time bin was then derived following a binned likelihood analysis (binned in space and energy), by performing a simultaneous fit of the source of interest and other {\it Fermi}-LAT sources. These sources, included in a $15\degr \times 15\degr$ region, were taken from the 3FGL catalog \citep{3fgl_catalog}, along with the Galactic and isotropic diffuse backgrounds (gll\_iem\_ext\_v06.fits and iso\_P8R2\_SOURCE\_V6\_v06.txt). We checked that the residual maps are well behaved (small fluctuations, $<3\sigma$).
For each source of interest, we first performed a likelihood fit over the entire time-interval data. The fit was carried out in an iterative way, in order to derive the best-fit values for the normalisation of all sources (point sources and diffuse components) in the region of interest (ROI). We also checked that each ROI was adequately modelled by inspecting the residual map and TS map of the ROI. New point-like excesses identified in the TS map were included in the ROI model when their significance was $TS>25$\footnote{This was done iteratively applying the fermipy tool \texttt{$find\_sources$}.}.

For the light curve bins, the diffuse components of the likelihood fit were fixed to the full-time interval average. Sources within 3 degree from the source of interest had their normalization left free to vary. Flux upper limits  were computed for all those time bins where the TS of the source of interest was lower than 4 ($\sim 2\sigma$).

\section{Methodology} \label{sec:methodology}
The solution we propose to manage our sample of almost 2300 AGN is to develop a periodicity-search pipeline, where each AGN is studied in the same systematic way. This pipeline is organized in different stages, which include data processing and decision making. In each of the processing stages, a specific set of algorithms for detecting evidence of periodicity is applied. Additionally, some techniques are used to infer the significance levels of the periods reported by such algorithms. These methods are presented in $\S$\ref{sec:methods}. Based on the above information, further constraints and selection criteria are defined to categorize an object as periodic-emission candidate. The entire pipeline is explained in $\S$\ref{sec:procedure}.

The main limitation in the periodicity search is the time-series noise. The identification of potential oscillations is complex since a large fraction of the variance is due to random fluctuations. These stochastic effects generally show larger amplitudes on longer timescales. In particular, this {\it red noise} impacts the lower frequencies since its spectral density is inversely proportional to frequency$^2$. Consequently, different methods are included in the periodicity-search pipeline to cope with this difficulty.

\subsection{Periodicity detection methods} \label{sec:methods}
In order to reduce biases, ten different algorithms are used in our methodology since all of them have drawbacks and advantages \citep{lomb_vdp, methods_critica_PKS_0735}.  To complement these algorithms, we use techniques to infer the significance of the periods provided by the search algorithms. The following subsections introduce briefly such algorithms and techniques.

\subsubsection{Lomb-Scargle}\label{sec:lsp}
The Lomb-Scargle periodogram (LSP) is one of the most employed and best known methods for detection of periodicity in time-series in astronomy, regardless the time-series contains evenly-spaced or unevenly-spaced data \citep{lomb_1976, scargle_1982}. LSP has been applied in several scenarios using slightly different methods and techniques \citep[e.g.,~][]{zhang_pks2155,ackermann_pg1553}.

In our pipeline, three different methods enable obtaining the significance of the peaks generated by LSP: power-law fitting, Bootstrap, and simulating LCs.

\paragraph {Power-law fitting}\label{sec:power_law}
This approach is a fast and simple test to calculate the significance of a LSP superposed on a red noise spectrum \citep{power_law}. This method assumes that the underlying noise continuum spectrum follows a power law. In particular, we use the algorithm implemented by \citet{power_law}, which is also used by \citet{zhang_pks2155} and \citet{sandrinelli_powerlaw}. 

\paragraph {Bootstrap}\label{sec:lsp_boostrap}
A more recent study contains a thorough analysis of the advantages and weakness of LSP, including different variants of this algorithm \citep{lomb_vdp}. One of them is the LSP with bootstrap, which is the most robust way to estimate the false alarm probability (FAP), which is an additional estimator of the significance of a peak in LSP \citep{prokhorov_set}. The FAP measures the probability that a dataset with no signal generates a peak of similar magnitude as a consequence of random fluctuations \citep{lomb_vdp}. In our pipeline, the periodogram is obtained with the Generalised Lomb-Scargle periodogram \citep[GLSP,][]{lomb_gen}. We use the algorithm presented by \citet{astroml}. 

\paragraph {Simulating light curves} \label{sec:methods_slc}
Another method to infer the significance of the LSP peaks is based on simulating LCs \citep{zhang_pks0301,zhang_pks0426}. In this work, we use the implementation of \citet{emma_lc}, coded in Python by \citet{connolly_code}. To obtain the significance, LCs are simulated, based on the best-fitting result of power spectral density and the probability density function of the original LC. For each simulated LC, a LSP is obtained. The confidence levels of the LSP peaks are calculated by using the percentiles of the power for each period bin in the LSPs of the simulated LCs.

\subsubsection{REDFIT}\label{sec:redfit}
REDFIT enables the calculation of the bias-corrected spectrum of a time-series, providing at the same time the significance of the peaks of such spectrum \citep{redfit}. This application estimates the red-noise spectrum from the data time-series, fitting a first-order autoregressive process. An autoregressive process is a mathematical model used to represent random time-varying processes \citep[e.g.,~][]{redfit}. 

In addition, REDFIT estimates the significance of the peaks in the spectrum of a time-series against the red-noise background \citep{gupta_redfit,sandrinelli_redfit}. REDFIT also provides the FAP levels of the peaks present in the periodogram (the maximum level it provides for the FAP is 2.5$\sigma$).

\subsubsection{Phase dispersion minimization}\label{sec:pdm}
The phase dispersion minimization (PDM) is an algorithm in which the data are placed into phase bins and the overall scatter within each bin is characterized with a parameter $\theta$ \citet{pdm}. Lower values of $\theta$ imply less scatter and, therefore, a better phasing \citep{tavani_pdm_pg_1553}. We use the implementation provided by \texttt{PyAstronomy} \citep{PyAstronomy}.   
In order to calculate the significance of the periods obtained by the PDM method, we use the technique presented by \citet{linnel_pdm}, which is based on Fisher' method of randomization. The technique is defined as follows: starting from the original time-series, the PDM is calculated and the $\theta$ of the lowest peak is determined. After that, the original time-series is randomly permuted and PDM is calculated again. If the lowest peak in the new PDM is lower than the original one, '1' is added to the count. This process is repeated a number of times where finally FAP is the value of the count of '1' divided by the number of times a permutation is performed in the original time-series \citep{tavani_pdm_pg_1553}. 

\subsubsection{Wavelet techniques}\label{sec:wavelets}
The first algorithm based on wavelets used in this work is the continuous wavelet transform \citep[CWT,][]{wavelet}. In the CWT analysis, we use the Morlet mother function. We use the implementation provided in \texttt{PyCWT}\footnote{\url{https://pypi.org/project/pycwt/}}, which also provides the significance of the peaks \citep[e.g.,~][]{espaillat_wavelet,ackermann_pg1553}.

Some AGN time-series are unevenly-spaced. To cover such scenario, we use the weighted wavelet Z-transform \citep[WWZ,][]{foster_wwz,zhang_pks0426,gupta_redfit}. To calculate the significance, we use the simulated light curves technique described in $\S$\ref{sec:lsp}. For the implementation, we use the Python package \texttt{Pyleoclim} \citep{Pyleoclim}. 

\subsubsection{Enhanced discrete Fourier transform}\label{sec:dft}
An additional algorithm employed for the periodicity detection is Welch's method \citep{welch_method, which} is based on the discrete Fourier transform (DFT). Compared to DFT that is sensitive to any noise, Welch's method is more robust to noisy data. This method splits the signal into segments, estimating the power spectral density for each segment, and averaging over these local estimates. The average reduces the variance of the estimated periodogram. This process trades some resolution in the frequency domain for improved robustness. Welch's method is complemented by using the Hanning window to reduce other spurious phenomena that can distort the detection \citep{methods_critica_PKS_0735}. 

For the implementation of DFT with Welch's method and Hanning window, we use the facilities provided by the Python package \texttt{SciPy} \citep{SciPy}. In order to get the significance of the peaks, we use Fisher’s method of randomization technique, previously explained in $\S$\ref{sec:pdm}.  

\subsubsection{Markov Chain Monte Carlo sinusoidal fitting}\label{sec:mcmc}
Another method used for the detection of the periodicity is to fit the LC to a sinusoidal \citep{emcee}. By means of Markov Chain Monte Carlo (MCMC), we fit the AGN LC according to the model,
\begin{equation} \label{eq:model_fitting}
\phi(t) = O + A\sin \bigg(\frac{2 \pi t}{T} + \theta \bigg).
\end{equation}

The parameters to be estimated are offset (O), amplitude (A), period (T) and phase ($\theta$). The results used in the periodicity analysis are the posteriors of each parameter. All the priors are constant distributions with values covering the following ranges:

\begin{itemize} 
	\item O: [0, 150] ($\times$ 10$^{-6}$ MeV cm$^{-2}$ s$^{-1}$)
	\item A: [0, 80] ($\times$ 10$^{-6}$ MeV cm$^{-2}$ s$^{-1}$)
	\item T: [0.5, 5] years
	\item $\theta$: [0, 360] degrees
\end{itemize}

For some sources, several MCMC sine fitting to the LCs are implemented in order to evaluate different potential periods. The comparison between these fittings are in terms of the likelihood ratio test (LRT), a statistic to evaluate which model fits better. Here, the LRT is represented as Test Statistic of Fitting (TS\textsubscript{Fitting}). We calculate TS\textsubscript{fitting} as,
\begin{equation}
TS\textsubscript{fitting}=-2[\ln \mathcal{L}(fitting 1) - \ln \mathcal{L} (fitting 2)]
\end{equation}

\noindent where $\mathcal{L}$ represents the likelihood, which is applied to two fitting hypothesis. 

In addition to TS\textsubscript{Fitting}, the difference in the degrees of freedom of the models are required to determine the statistical significance of the difference between the models. Finally, the LRT statistic approximately follows a chi-square distribution. Therefore, using TS\textsubscript{Fitting} and the degrees of freedom, we obtain the p-value of the model comparison. 
\subsubsection{Bayesian quasi-periodic oscillation}\label{sec:bayesian}
Another method to study the impact of red noise is to search for periodicity using a Bayesian approach for quasi-periodic oscillation detection \citep[Bayesian QPO,][]{huppenkothen_bayesian}. The method basically compares two noise models, a simple model that acts as null hypothesis and a more complex model as alternative hypothesis. The simple model is a power-law since, as explained in $\S$\ref{sec:lsp}, the red noise spectrum has approximately a power-law behavior. The second model is a bent power law \citep[broken power law with smooth transition,][]{huppenkothen_bayesian}. The LC is fitted to both models, obtaining a LRT (specifically known as B-LRT) to quantify how unlikely or likely the LC are generated from the simple model. By means of a large number of simulated periodograms from a MCMC sample, a distribution of LRTs from the simple model is created after fitting these periodograms by both noise models. From here, we obtain the tail-area probability (p-value) of LRT. 

Furthermore, using the result from the previous model comparison, this method allows the detection of periodicity by binning the periodogram. The rationale behind this is that the periodic oscillations can be narrow or broad: a single oscillation may be spread out over several frequency bins, or it may be so coherent that it is mostly concentrated in one bin. In this latter case, sampling will usually cause the oscillation to be spread over two adjacent bins. Then, we pick the frequency bins with the largest maximum likelihood from a MCMC sample. For each frequency bin, the method provides the sensitivity, specifically, the fractional root mean square amplitude at periods defined by the user.

\begin{figure*}
\includegraphics[width=2\columnwidth]{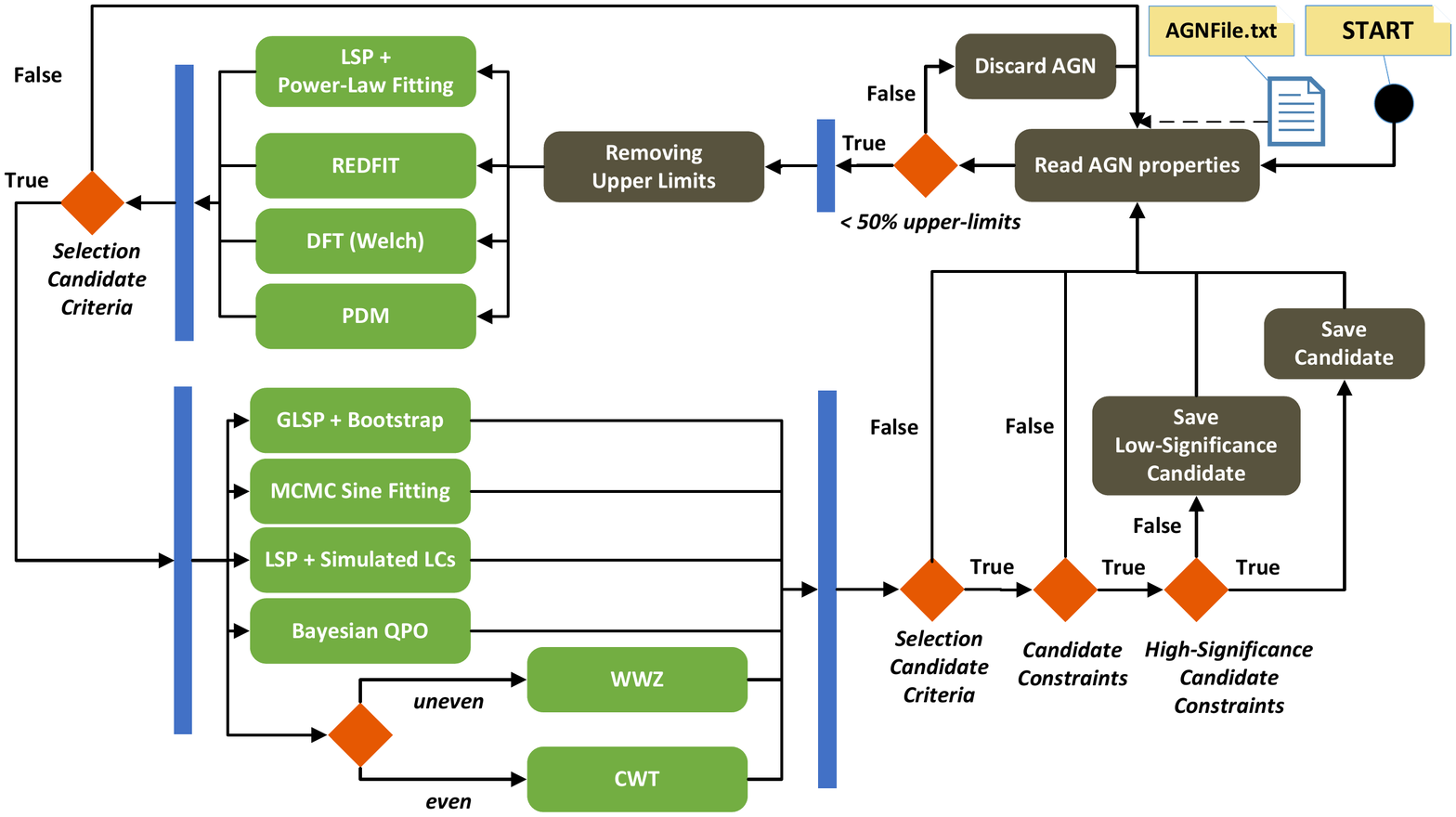}
\caption{Periodicity-search pipeline summarized in an UML activity diagram.}
\label{fig:study_flow}
\end{figure*}

\subsection{Periodicity-search pipeline} \label{sec:procedure}
To search for periodicities in our AGN sample, we create a periodicity-search pipeline. This pipeline is built using a hierarchical structure, composed by different processing and decision stages. In each processing stage, we apply a set of algorithms previously presented, according to their functional characteristics. The decision stages are defined by a set of constraints and selection criteria, related to the properties of the LC to be processed and how significant the detection of periodic emission is. The structure of the periodicity-search pipeline is shown in Figure~\ref{fig:study_flow}. The specification of the pipeline shown in this figure is implemented according to the standard Unified Modeling Language (UML), captured in an activity diagram\footnote{for more details, see \url{http://www.uml.org}}.

\paragraph {Filtering based on upper-limit energy fluxes}\label{sec:filter_upli}
The pipeline starts reading all the relevant information required by the periodicity study (type, date, energy flux, energy flux uncertainty). We do not use the upper limits in the periodicity analysis process.

As starting point of the analysis, the LCs are checked in relation to the fraction of bins with upper-limit energy fluxes, LCs with more than 50\% of upper limits are rejected (see Figure~\ref{fig:study_flow} and $\S$\ref{sec:upper-limits}). After this filtering, the remaining sample contains 351 AGN (15\% of the initial sample).

\paragraph {Coarse analysis}\label{sec:fast_analysis}
Now, we apply the first group of algorithms. The initial group of methods is characterized by requiring less computation time, enabling a fast periodicity characterization. These methods include LSP + power-law fitting ($\S$\ref{sec:lsp}), REDFIT ($\S$\ref{sec:redfit}), DFT (with Welch's method, $\S$\ref{sec:dft}) and PDM ($\S$\ref{sec:pdm}). In order to obtain the significance of the peaks detected in the periodograms, we use the following parameters:

\begin{itemize}
	\item REDFIT: 10,000 Monte Carlo simulations. 
	\item DFT (with Welch's Method): 20000 permutations in Fisher’s method of randomization.
	\item PDM: 10,000 permutations in Fisher’s method of randomization.
\end{itemize}

\begin{figure}
\includegraphics[width=\columnwidth]{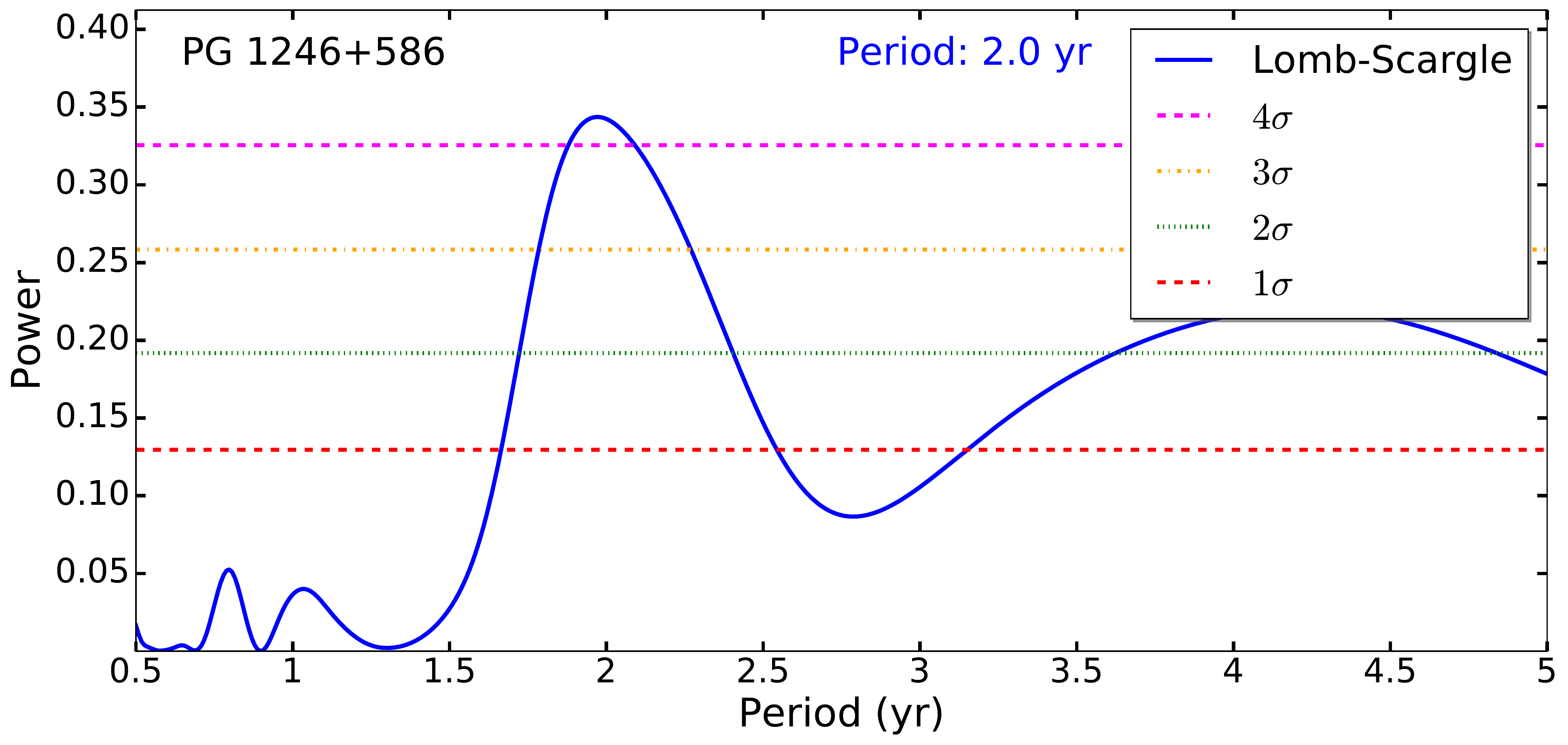}
\caption{Example of the periodogram for PG 1246+586  using the results from GLSP + bootstrap.}
\label{fig:periodogram}
\end{figure}

We have to define some criteria to categorize an object as candidate to emit periodically (see Figure~\ref{fig:study_flow}). The criteria of periodic-emission candidate selection in these stages is a combination of a two-step filter: (i) the corresponding periodogram generated by each algorithm must have a peak above a ``loose'' significance level L1 (see Figure~\ref{fig:periodogram}); (ii) at least one periodogram must have one peak above a ``tight'' significance level L2. These L1 and L2 levels are specific for each algorithm. These significance thresholds were selected in order to keep the contamination of spurious periodicity detection (explained in $\S$\ref{sec:fpdr_results}) under 0.5\%. This criteria is captured in Figure \ref{fig:study_flow} by the tag ``Selection Candidate Criteria''. By being flexible in such decision stage, we want to avoid losing potential periodicity candidates.

For the periodogram generated by each algorithm, the loose level of significance is 1$\sigma$. Regarding a tighter level of significance, the chosen value is $\geq$2$\sigma$ for LSP + power-law fitting ($\S$\ref{sec:lsp}), DFT with Welch's method ($\S$\ref{sec:dft}), PDM ($\S$\ref{sec:pdm}) and REDFIT ($\S$\ref{sec:redfit}). After applying this initial and fast search, the remaining sample contains 98 AGN (4\% of the initial sample).

\paragraph {Fine Analysis}\label{sec:fine_analysis}
This subsample of 98 AGN is fed into the next analysis stage, composed of the rest of the methods, those which require more computational power\footnote{We compared the results from these ``coarse''/fast analysis methods with those from the ``fine''/slow analysis methods on a random subsample of AGN and the results were similar, concluding that there is no bias between these two sets of methods.}: GLSP + bootstrap ($\S$\ref{sec:lsp}), LSP + simulated LC ($\S$\ref{sec:lsp}), and MCMC sine fitting ($\S$\ref{sec:mcmc}).

For the wavelet algorithms, we need to distinguish between the LC type; as a consequence of removing the upper-limits, some of the LCs become uneven time series (irregularly distributed). Therefore, we define two different branches for evenly or unevenly-spaced LCs. For the even LCs, the method considered is the CWT ($\S$\ref{sec:wavelets}). For the uneven LCs, this former method is replaced by the WWZ ($\S$\ref{sec:wavelets}). In order to compute the significance of the peaks detected, we used the following parameters:

\begin{itemize}
	\item GLSP + bootstrap: 10,000 resamplings. 
	\item LSP + simulated LC: we simulate 15,000 LCs for each AGN, using 1000 iterations for the fitting of the original LC.
	\item MCMC sine fitting: to perform the parameter estimation we use 100 walkers, 20,000 iterations and 3000 ``burn-in'' steps to enable the stabilization of the MCMC.
	\item WWZ: In this case and due to the long computation time required for each WWZ process iteration, we use 3000 simulated LCs with 1000 iterations for the fitting of the original LC.
\end{itemize}

With this second group of algorithms, we use the same constraint corresponding to the loose level of significance. The constraint on the tight level of significance of the peak (or peaks) is $\geq$2$\sigma$, due the same reason previously mentioned. 
When the period we obtain is incompatible with the period found by a previous work, the LC is plotted along with the sine reconstruction from these two different periods. Then, we use a likelihood ratio test to compare statistically both results.

Complementarily, we also apply the Bayesian QPO method at this second stage, obtaining a probability of the influence of red noise in the LC analysis. In order to perform the Bayesian analysis, we use 10,000 simulated periodograms with 10,000 MCMC iterations and 200 walkers. For B-LRT $\leq$5\%, the red noise hypothesis is rejected. For higher B-LRT, it means that the period detection may be produced by red noise. Additionally, this method provides the residuals for both noise models. Strong peaks in the residuals indicate evidence of periodicity. 
Furthermore, for the Bayesian QPO detection, we use the same previous MCMC configuration and select the objects with a p-value of $\leq$5\%, in at least two or more bins. Then, we represent the sensitivity of the set of periods (specifically, 100 points in the range 0.5--5.5~yr) in two frequency bins ($\S$\ref{sec:bayesian}), finding the period with the highest sensitivity in each bin. 
Combining all the aforementioned methods, constraints, and criteria, the number of sources that remains is 65 (3\% of the initial sample).

We filter these sources by imposing a new condition: at least, three methods must provide a detection at $\geq$3$\sigma$ at the same period (we note that for REDFIT, the significance is $\geq 2.5\sigma$, which is the maximum allowed by the method). This constraint is captured in Figure~\ref{fig:study_flow} by the tag ``Candidate Constraints''. There is, however, an exception to this selection criterion. This exception includes two situations (1) when an algorithm does not find a compatible period and (2) when a compatible period is found with low significance (in terms of the tight level). 

Finally, to select the highly significant periodicity candidates we impose a last constraint: at least, four methods must provide $\geq$4$\sigma$ at the same period. This condition is captured by the tag ``High-Significance Candidate Constraints'' in Figure~\ref{fig:study_flow}. With this criterion the contamination of spurious periodicity detection is $<$ 0.5\% (see $\S$\ref{sec:fpdr_results}).

\section{Results and discussion} \label{sec:candidates}
\begin{figure*}
\includegraphics[width=2\columnwidth]{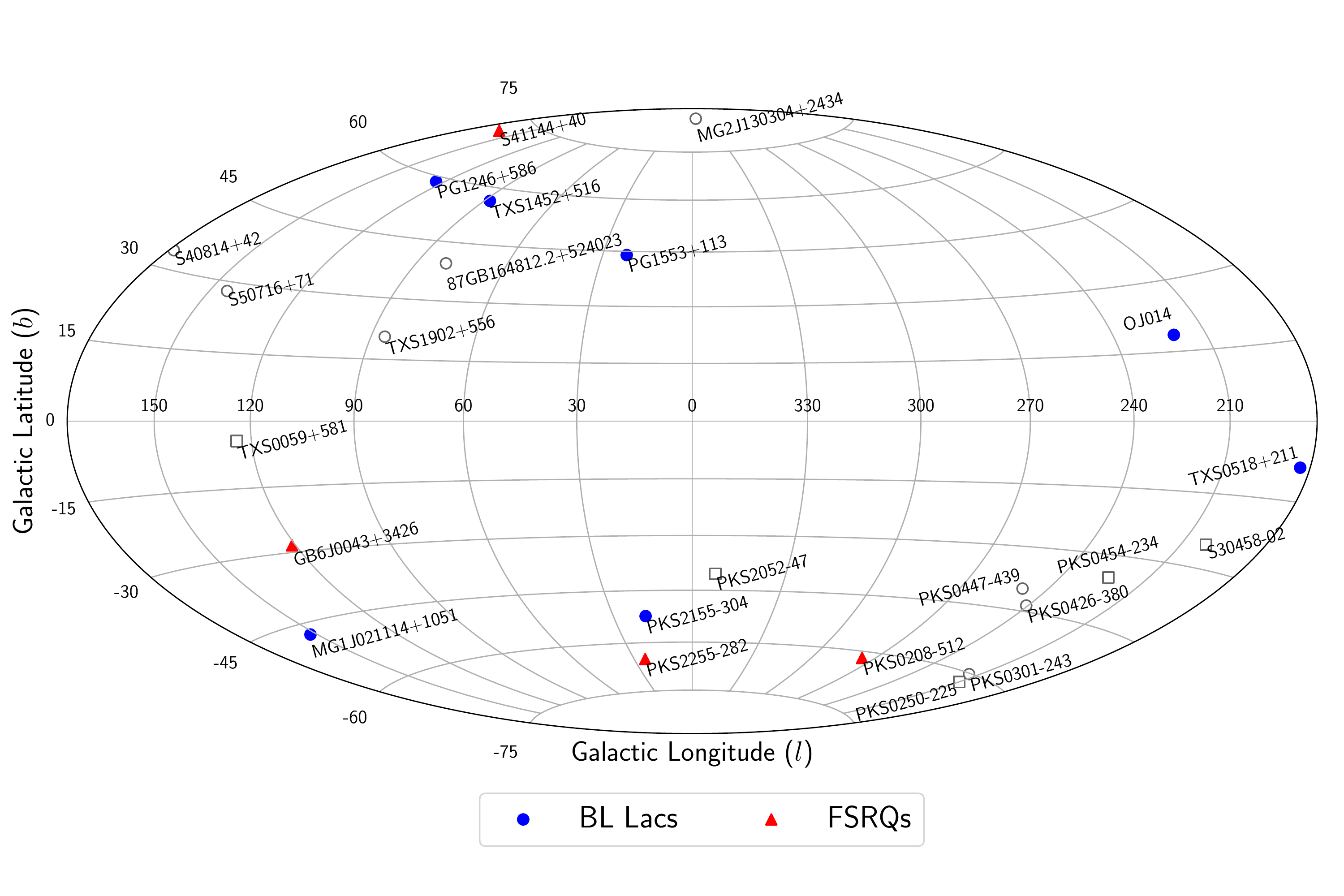}
\caption{Sky map showing the 11 sources with periodic emission (filled colored symbols) plus 13 with lower significance (open gray symbols) tagged by their association name. These sources are represented with different symbols according to their source type, BL Lacs (circles) and FSRQs (triangles). Galactic coordinates and Hammer-Aitoff projection are used. The spatial inhomogeneity in the detection of the periodic-emission candidates is produced from different exposures by {\it Fermi}-LAT.}
\label{fig:skymap}
\end{figure*}

From the periodicity-search pipeline, we identify 11 AGN with evidence of periodic $\gamma$-ray emissions. Table \ref{tab:candidates_list} lists these 11 periodic-emission candidates and Figure~\ref{fig:skymap} shows their location on the sky. According to their AGN type, most of them (7) are BL Lacs and 4 FSRQs \citep[these source identifications are taken from][i.e.,~4FGL]{4fgl_2019}. Furthermore, most of the candidates are at moderate/high redshifts, which may support the idea of binary SMBH systems as explanation for their periodic behavior \citep[see e.g.~][]{rieger_2007,prokhorov_set}.

Our candidates for having periodic emission are presented in the next subsections. $\S$\ref{sec:sub_candidates_a} includes the candidates whose periodic behaviour were previously reported in the literature (denoted in Tables \ref{tab:candidates_list} by a star). New candidates are shown in $\S$\ref{sec:sub_candidates_b}. In $\S$\ref{sec:others}, we present sources that have high significance from some of the methods but do not satisfy the last of our conditions ($\S$\ref{sec:procedure}) to be considered as highly significant candidate. These other sources are presented in bottom section of Table~\ref{tab:candidates_list}.

\subsection{Candidates in the literature} \label{sec:sub_candidates_a} 
Our 11 periodic-emission candidate sample includes 2 AGN previously reported in the literature as having periodic behavior (see Table \ref{tab:a_candiadtes_periods}). These are PG~1553+113 and PKS~2155$-$304.

PG 1553+113: \citet{ackermann_pg1553}, \citet{tavani_pdm_pg_1553}, \citet{prokhorov_set} and \citet{sandrinelli_powerlaw} find a period of $\sim$2.2~yr with high significance, compatible with the result found in the present work of $\sim$2.2~yr. \citet{covino_negation} finds no periodic $\gamma$-ray emission for this object.

PKS 2155$-$304: This source has been found to be periodic by \citet{sandrinelli_powerlaw}, \citet{zhang_pks2155} and \citet{prokhorov_set}, with periods of 1.70~yr, 1.74~yr and 1.76~yr, respectively. These results are compatible with our period of $\sim$1.7~yr. Once again, \citet{covino_negation} claim the absence of any periodic $\gamma$-ray emission in this object.

\subsection{New periodic-emission candidates} \label{sec:sub_candidates_b}
We find 9 sources not previously identified in the literature for having periodic emissions listed in Table \ref{tab:a_candiadtes_periods}. 

The period inferred for the object OJ~014 is $\sim$4.3~yr, which is close to the limit of the peaks to be detected according with the time interval of the data ($\sim$9~yr). As explained in $\S$\ref{sec:methodology}, the red noise has a larger impact in these long period ranges \citep[short frequencies,][]{vaughan_negation}. Table~\ref{tab:a_candidates_ts} shows the parameter B-LRT, which is related to the red noise impact \citep[$>5\%$ is interpreted as the source possibly being dominated by red noise,][]{huppenkothen_bayesian}.

\begin{figure}
\includegraphics[width=\columnwidth]{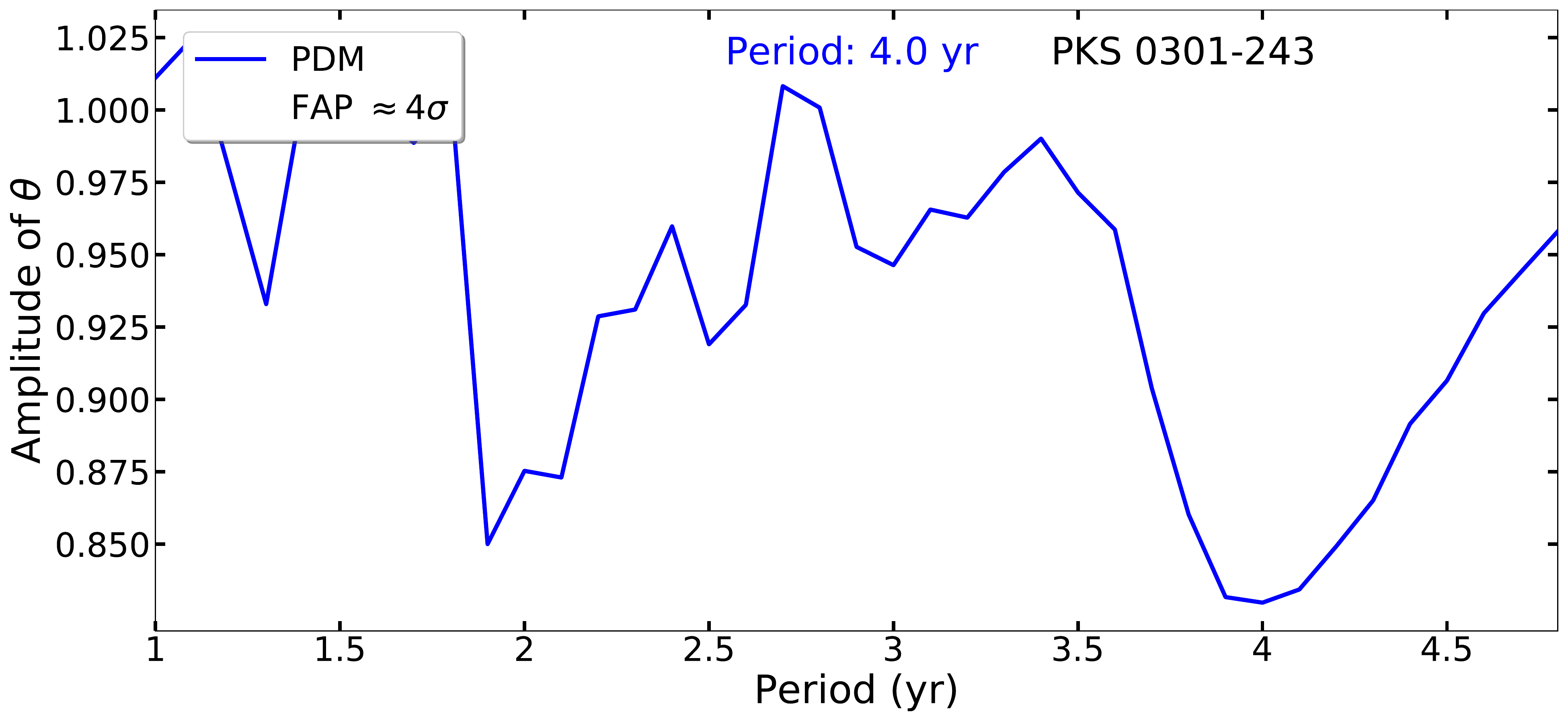}
\includegraphics[width=\columnwidth]{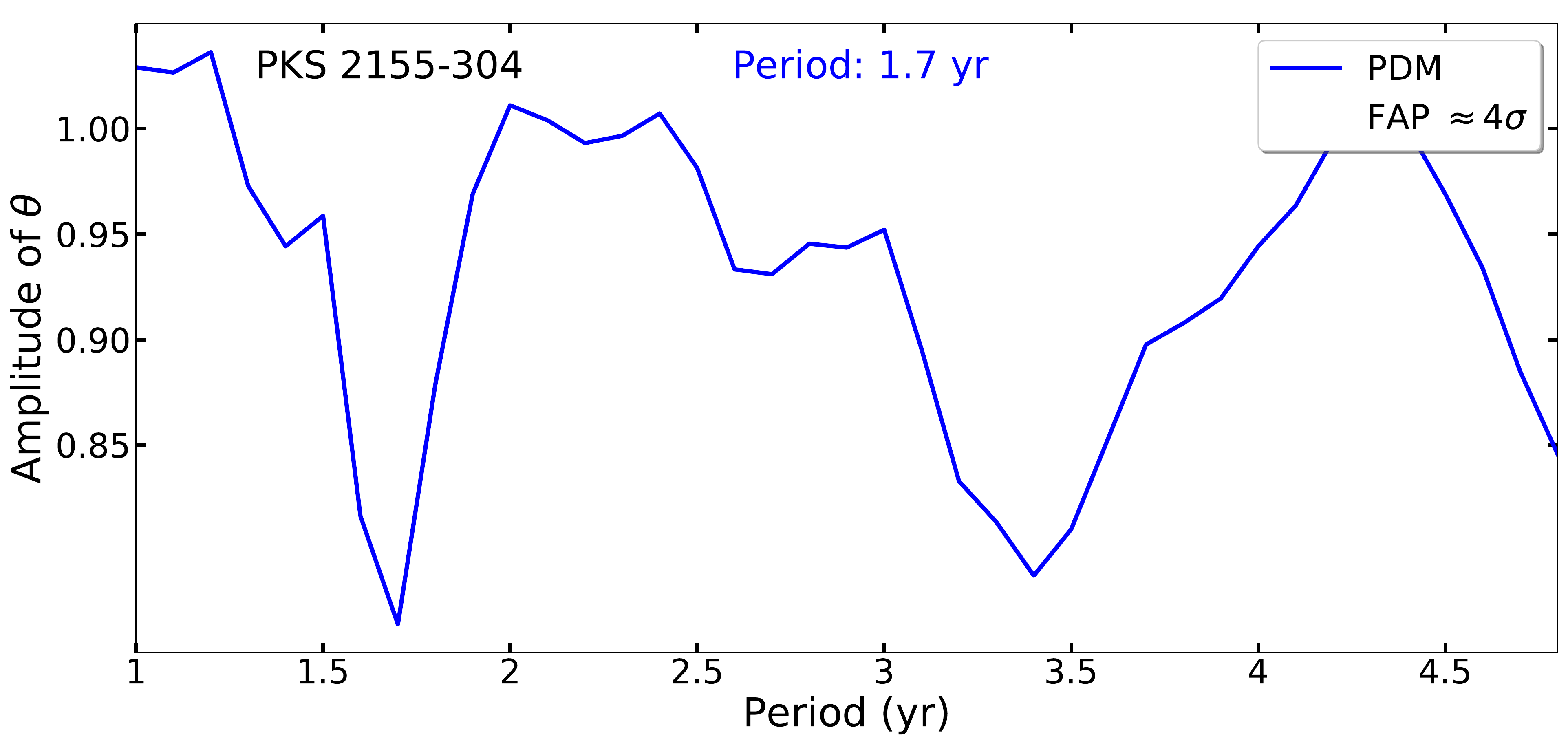}
\caption{Examples of results produced by the PDM method. {\it Top:} PKS 0301$-$243 with compatible period of $\sim$2 yr. {\it Bottom:} PKS 2155$-$304 with compatible period of $\sim$1.7~yr.}
\label{fig:pdm_problem}
\end{figure}

Looking at the results provided by the PDM method, we find some cases with large minima in the harmonics. In these cases, when the harmonic is closer to the limit of the period's detection (given by half of the total exposure, thus $\sim$4.5~yr in our case), the harmonic tends to have larger amplitude than the main, placed on the period. This effect may produce rather different periods from different methods (see Figure \ref{fig:pdm_problem}). However, when this harmonic is further from the detection limit, the result tends to be compatible with those from our other methods (see Figure \ref{fig:pdm_problem}). The sources with large minima in the harmonics are marked in Table~\ref{tab:a_candiadtes_periods}.

In general, the periods reported by the Bayesian-QPO method are coherent with the other methods (PKS 2255$-$282 is the exception). 
In order to denote the presence of flares, we define a selection criterion to detect these high-activity phenomena in the LCs. We use the results provided by the MCMC sine fitting, this is, the offset and the amplitude parameters (see Equation \ref{eq:model_fitting}). Then, for each periodicity candidate, we use as reference level \textit{$\sim$3$\times$(offset + amplitude)}, marked as ``Flare'' in Table~\ref{tab:a_candidates_ts}.

We can estimate the necessary exposure to get $\geq 5\sigma$ in the detection of periodicity for these 11 sources. The procedure is adding cycles at the end of our observations assuring the continuity of the LCs. These cycles are taken by visual inspection from each LC. For each new LC, we apply the methods presented in $\S$\ref{sec:procedure} (except WWZ due to computational limitations and REDFIT because there a confidence limit of $2.5\sigma$). In this way, we estimate the number of cycles necessary for a $5\sigma$ detection (see Table \ref{tab:estimation_5sigma}). This table also shows how these cycles translate to years of LAT exposure.

\subsection{Low significance candidates} \label{sec:others}
During the analysis, several objects present evidence of periodical $\gamma$-ray emission near the limit of our criteria. However, we think they may deserve future attention when more data is available. All these sources are filtered at the last decision stage in Figure \ref{fig:study_flow} (see $\S$\ref{sec:procedure}). 

This subsample includes 4 AGN previously reported in the literature as having periodic behavior (see bottom section of Table \ref{tab:candidates_list});  these are PKS~0301$-$243, PKS~0426$-$380,  S5~0716+71, and PKS 2052-47.

PKS 0301$-$243: \citet{zhang_pks0301} conclude that this source has a period of 2.1~yr, which is similar to our result of $\sim$2.1~yr. However, \citet{covino_negation} claim there is no evidence of periodicity in this object.

PKS 0426$-$380: \citet{zhang_pks0426} obtain a period of 3.3~yr compatible with ours of $\sim$3.1~yr. \citet{covino_negation} also disagree with this periodicity detection. 

PKS 2052$-$47: \citet{prokhorov_set} obtain a period of 1.7 yr compatible with ours of $\sim$1.7 yr.

\begin{figure}
\includegraphics[width=\columnwidth]{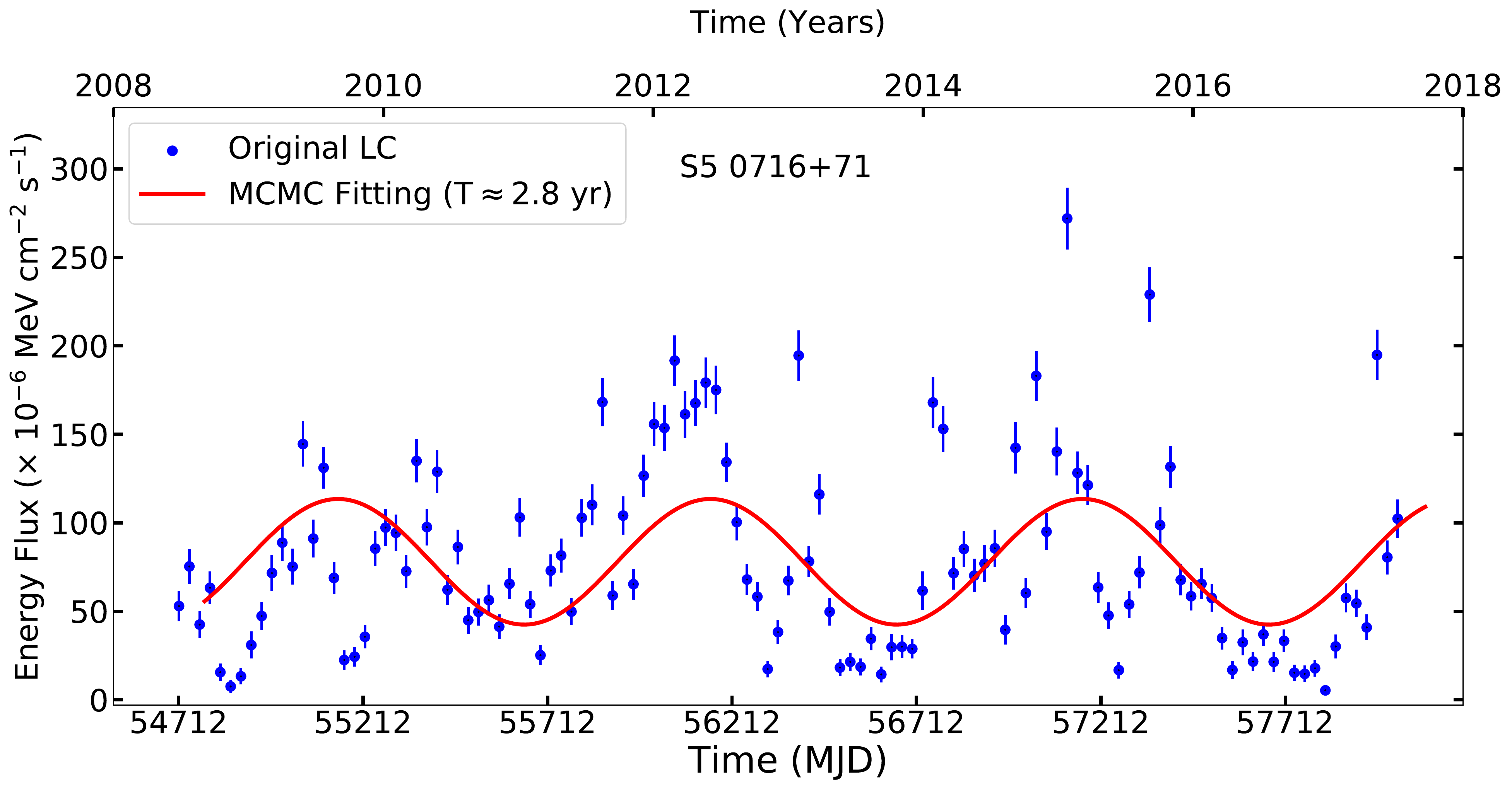}
\includegraphics[width=\columnwidth]{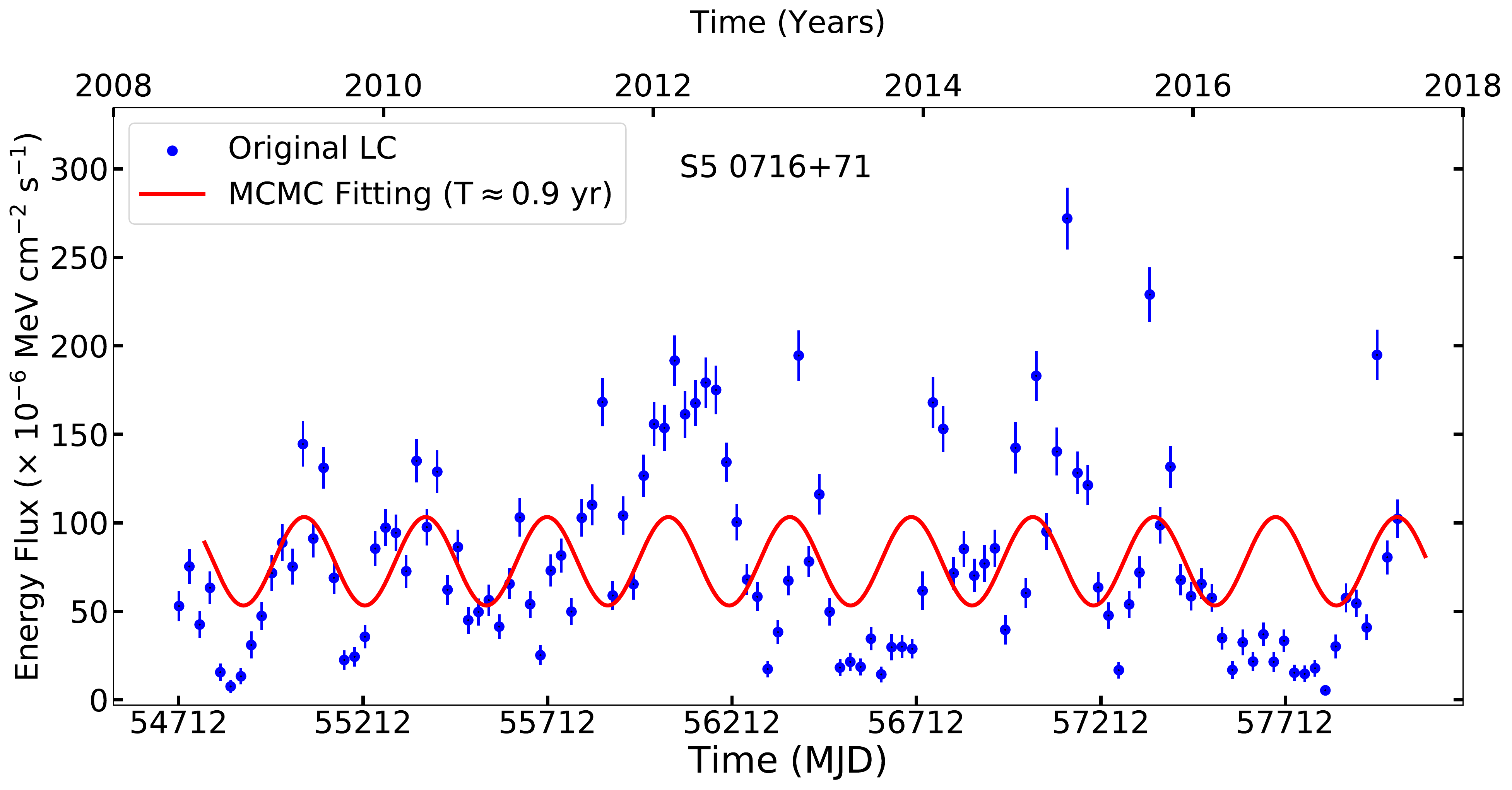}
\includegraphics[width=\columnwidth]{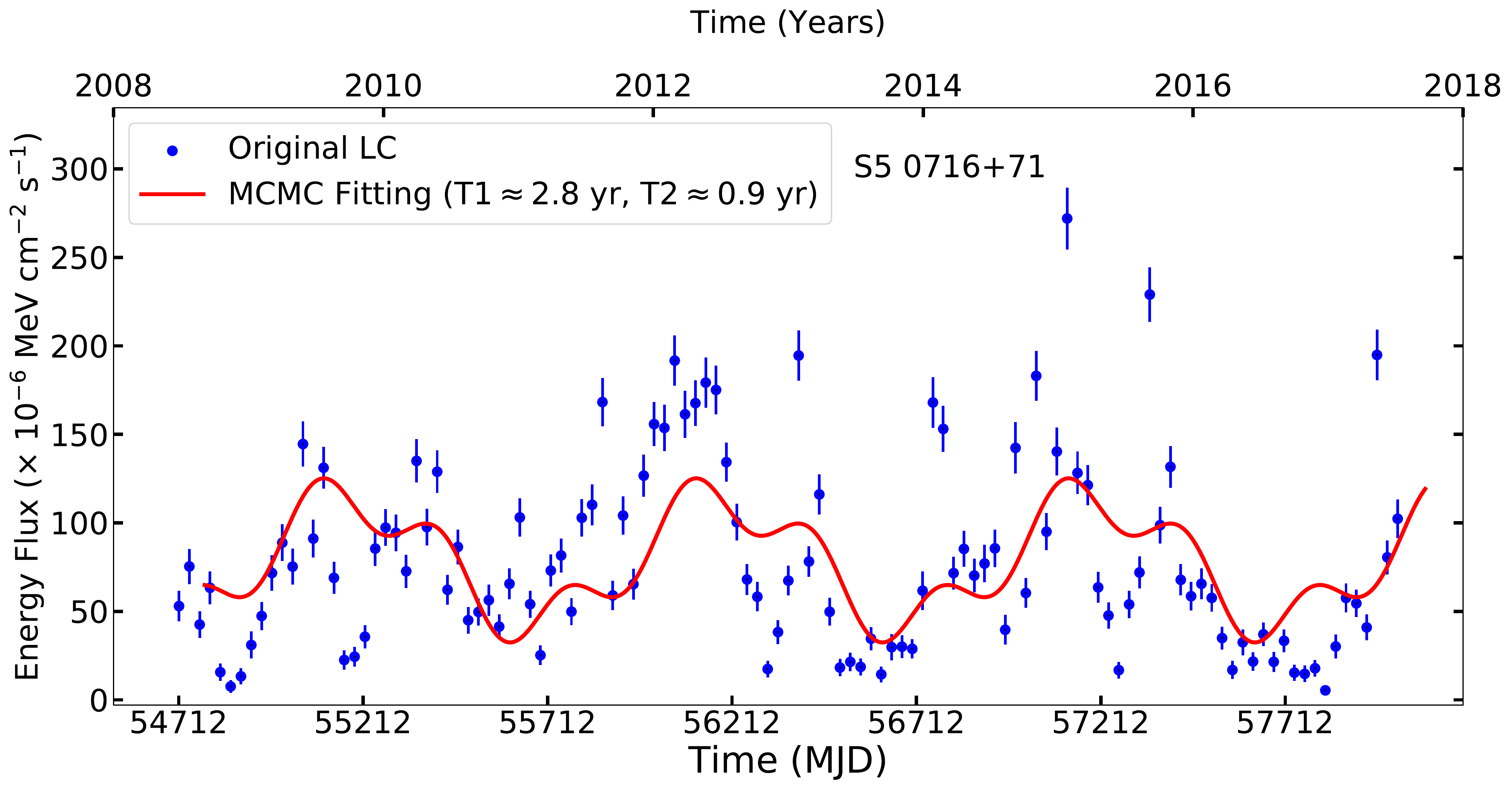}
\caption{MCMC sine fitting reconstructions of S5 0716+71 for different periods:	top, T$\sim$2.7~yr; middle, T$\sim$0.9~yr; bottom, T1$\sim$2.7~yr and T2$\sim$0.9~yr.}
\label{fig:s5_0716}
\end{figure}

S5 0716$+$71: \citet{prokhorov_set}, \citet{sandrinelli_S5_0716_71} and \citet{li_S5_0716_714} find periodic emission around 345 days ($\sim$0.9 yr), however we obtain the most significant period at $\sim$2.8~yr (nearly a multiple of 0.9 yr). Interestingly, using some of our methods, we find a peak located $\sim$1~yr, which is compatible with the quoted value of $\sim$0.9~yr. We evaluate both scenarios by plotting the results from a MCMC sine fitting (as explained in $\S$\ref{sec:procedure}). Both sine reconstructions are shown in Figure~\ref{fig:s5_0716}. The value of the TS\textsubscript{fitting} defined in $\S$\ref{sec:mcmc} is $\sim5.7~\sigma$, which implies that the $\sim$2.8~yr fit is better.

Additionally, we perform a MCMC sinusoidal fitting considering two sine components according to the equation,

\begin{equation} \label{eq:model_fitting_2}
\phi(t) = O + A_{1}\sin \bigg(\frac{2\pi t}{T_{1}} + \theta_{1}\bigg) + A_{2}\sin \bigg(\frac{2\pi t}{T_{2}} + \theta_{2}\bigg)
\end{equation}

The periods (i.e., the variables \textit{T1} and \textit{T2}) are constrained in the fits using $T1=0.9^{+0.6}_{-0.1}$~yr and $T2=2.8\pm0.1$~yr. We compare the best fit of the double-sine scenario against the two single-sine scenarios using the LRT.
For both sine models comparisons, the TS\textsubscript{fitting} is greater than the value of the chi-square distribution for three degrees of freedom (difference between the number of fitting parameters of both tested models) and a p-value of 0.05. The double-sine scenario is thus not statistically preferred.

In general, the periods reported by the Bayesian-QPO method are coherent with the other methods. However, in some of them the results are not compatible (PKS 0454$-$234, S3 0458$-$02, S5 0716+71, 3FGL J1649.4+5238).

Recently, \citet{bhatta_mrk501} claimed that the $\gamma$-ray emissions of Mrk 501 presents a periodicity of $\sim$1~yr with a weekly binning of about ten years of {\it Fermi}-LAT data. According to our analysis (with nine years of data and 28-day binning), no evidence of periodic emission was reported by our analysis pipeline. Similarly, \citet{bhatta_mrk421} claim that the $\gamma$-ray emission of Mrk 421 has a period of $\sim$1~yr. Our analysis does not confirm this finding.

OJ~287 was studied by \citet{sandrinelli_redfit}, estimating a period of $\sim$1.1~yr, which is compatible with the period we obtained, $\sim$1.1~yr, but the significance tends to be lower than $2\sigma$. Considering these results, we agree with \citet{goyal_no_oj287}, who do not find any periodicity. 

For BL Lacertae, \citet{prokhorov_set} and \citet{sandrinelli_S5_0716_71} obtain a period of $\sim$1.9~yr and $\sim$1.8~yr, respectively. According to our results, the period inferred by our different methods is $\sim$4.5~yr with low significance (lower than $2\sigma$), thus we do not find any periodicity. 

\subsection{Uncertainty quantification}
We perform two complementary analyses in order to evaluate our results, first, estimating the false-positive detection rate and second, checking the effect of upper limits on the results.

\subsubsection{False-positive detection rate} \label{sec:fpdr_results}
Given the large number of sources in our analysis, we need to calculate the false-positive detection rate (FPDR). This computation tells us about the possible contamination of our results due to stochastic effects. We use the $\sim$250 LCs that got rejected by the ``Coarse analysis'' ($\S$\ref{sec:procedure}) since they do not show periodic behaviour. We apply the method described in $\S$\ref{sec:lsp}, generating 120 simulated LCs for each source. Therefore, we obtain a new sample of $\sim$30,000 LCs. Then, our methodology is repeated for all methods except WWZ due to computational limitations. In summary:
\begin{itemize}
	\item GLSP + Bootstrap: 2000 resamplings. 
	\item LSP + Simulated LC: we simulate 500 LCs for each rejected AGN, using 100 iterations for the fitting of the original LC.
	\item REDFIT: 2000 MC simulations. 
	\item DFT (with Welch's Method): 5000 permutations for Fisher’s Method of Randomization.
	\item PDM: 500 permutations for Fisher’s Method of Randomization.
\end{itemize}

Next, we use the same selection criteria defined in $\S$\ref{sec:procedure} to identify periodic-emission candidates. As result, we obtained 31 spurious periodicity candidates in 29,000 LCs (from some rejected LCs it was not possible to generate all the simulated LCs; the parameters of the power spectral density and the probability density function were not obtained). The FPDR is the ratio \textit{number of candidates/number of LCs simulated} obtaining a $0.02\%$. Applying this FPDR to our original AGN sample of 2274 objects results in 1 periodicity candidate, which may be a spurious detection. 

\subsubsection{The impact in the results of upper limits in LCs} \label{sec:upper-limits}

The first filter that we applied to our original AGN sample was to remove all those with LCs with more than 50\% of upper limits. In this section, we use PG~1553+113 to evaluate the impact in the period detection of the existence of upper limits in the LCs. PG~1553+113 is used because it is detected in each time bin and also it features the most significant periodicity. The procedure is the following. First, we use the median of the energy flux as reference level to create {\it fake} upper limits under this median. Then, we remove a percentage of them, ranging from 10\% to 70\% in increments of 10\%. This exercise is repeated 100 times for each percentage. The new LCs are analyzed by the methods employed in $\S$\ref{sec:fpdr_results} using all same parameters. To calculate the loss of significance, we consider the period of 2.2~yr and the significance level of $>$4$\sigma$ as reference values.

\paragraph{Power-Law Fitting}
After removing 10\% of the data, the loss of significance is $\sim$20\%, being $\sim$25\% after removing 50\% of the data. For the period, the value is shifted $\pm 5\%$ from the reference. Removing 60\% of data leads to a loss of significance of 40\%. In the case of removing 70\% of the data, the period is shifted $\pm 10\%$ and the period looses a 50\% of significance.

\paragraph{GLSP + Bootstrap}
In this case, no significant loss of sensitivity is detected until 50\% of the data have been removed. At that point, we register a loss of $\sim$40\% in the significance and the period is shifted by $\pm 5\%$. Removing 70\%, the  loss of significance is 60\% and the period shift is $\pm 10\%$.

\paragraph{LSP + Simulated LC}
This method presents similar results to the previous one.

\paragraph{DFT (with Welch's Method)}
When removing 50\% of the data, the significance decreases by 50\% and the period is shifted $\pm 40\%$. By removing 70\% of the data, the significance is reduced by 60\%.

\paragraph{PDM}
Typically, for PDM, the loss of significance is about 20\% when removing 50\% of the data. This loss increases up to 50\% when removing 70\% of the data.

\paragraph{REDFIT}
This method seems rather stable relative to removing data. For the case of removing 50\% of the data, the significance is reduced only by 5\%.

\section{Summary} \label{sec:conclusion}
In this work, we have implemented a systematic search for detecting periodical $\gamma$-ray emission from 2274 AGN detected with {\it Fermi}-LAT over the first 9 years of data. We design and develop a periodicity-search pipeline composed of ten different period-detection algorithms that are widely employed in the literature. These algorithms are complemented with a set of techniques to obtain the significance level of potential periods. The number of candidates with high-significant evidence of periodic $\gamma$-ray emissions is 11 (4 FSRQs and 7 BL Lacs). Out of these 11 sources, there are 9 whose periodic behavior has not been previously identified. Additionally, we find other 13 sources with low-significance periodicity. From these 13 sources, 9 of them have not being previously identify as potential periodicity candidates. This is the first large sample of $\gamma$-ray periodic emitters that has been ever found, which will further the study of properties of this type of sources and the understanding of their astrophysical nature.

\section*{Acknowledgements}
P.P., A.D., and J.A.B. acknowledge the support of the FPA2017-85668-P of the Agencia Estatal de Investigación del Miniserio de Ciencias, Innovación y Universidades. A.D. is also thankful for the support of the Ram{\'o}n y Cajal program from the Spanish MINECO. We thank the anonymous referee for their careful review.

The \textit{Fermi} LAT Collaboration acknowledges generous ongoing support from a number of agencies and institutes that have supported both the development and the operation of the LAT as well as scientific data analysis. These include the National Aeronautics and Space Administration and the Department of Energy in the United States, the Commissariat \`a l'Energie Atomique and the Centre National de la Recherche Scientifique / Institut National de Physique Nucl\'eaire et de Physique des Particules in France, the Agenzia Spaziale Italiana and the Istituto Nazionale di Fisica Nucleare in Italy, the Ministry of Education, Culture, Sports, Science and Technology (MEXT), High Energy Accelerator Research Organization (KEK) and Japan Aerospace Exploration Agency (JAXA) in Japan, and the K.~A.~Wallenberg Foundation, the Swedish Research Council and the Swedish National Space Board in Sweden. Additional support for science analysis during the operations phase is gratefully acknowledged from the Istituto Nazionale di Astrofisica in Italy and the Centre National d'\'Etudes Spatiales in France. This work performed in part under DOE Contract DE-AC02-76SF00515.

\software{
      astroML \citep{astroml},
      emcee \citep {emcee}, 
      fermipy software package \citep{Wood:2017yyb},
      PyAstronomy \citep{PyAstronomy},
	  PyCWT, 
	  Pyleoclim \citep{Pyleoclim},
      REDFIT \citep{redfit},
      SciPy \citep {SciPy},
	  Simulating light curves \citep{connolly_code}
}

\begin{deluxetable*}{cccccccc}[h]
	\tablecaption{{\it Top:} list of the 11 periodic-emission candidates, with their \textit{Fermi}-LAT name, coordinates, AGN type, redshift, association name and period (in years) obtained with the periodicity-search pipeline. {\it Bottom:} list of 13 AGN with low significance period detection. 
	\label{tab:candidates_list}} 
	\tablewidth{0pt}
	\tablehead{
		\colhead{3FGL Source Name} &
		\colhead{RAJ2000} &
		\colhead{DecJ2000} &
		\colhead{Type} &
		\colhead{Redshift} & 
		\colhead{Association Name} & 
		\colhead{Period}
	}
	\startdata        
	J0043.8+3425 & 10.96782 & 34.42687 & fsrq & 0.966 & GB6 J0043+3426 & 1.8 \\
	J0210.7$-$5101 & 32.68952 & $-$51.01695 & fsrq & 1.003 & PKS 0208$-$512 & 2.6 \\	
	J0211.2+1051 & 32.81532 & 10.85811 & bll & 0.2 & MG1 J021114+1051 & 1.7 \\ 
	J0521.7+2113 & 80.44379 & 21.21369 & bll & 0.108 & TXS 0518+211 & 2.8 \\ 
	J0811.3+0146 & 122.86418 & 1.77344 & bll & 1.148 & OJ 014 & 4.3 \\
	J1146.8+3958 & 176.73987 & 39.96861 & fsrq & 1.089 & S4 1144+40 & 3.3 \\
	J1248.2+5820 & 192.07728 & 58.34622 & bll & -- & PG 1246+586 & 2.2 \\
	J1454.5+5124 & 238.93169 & 11.18768 & bll & -- & TXS 1452+516 & 2.1 \\
	J1555.7+1111 & 238.93169 & 11.18768 & bll & 0.36 & PG 1553+113 & 2.2 \\
	J2158.8$-$3013 & 329.71409 & $-$30.22556 & bll & 0.116 & PKS 2155$-$304 & 1.7 \\ 
	J2258.0$-$2759 & 344.50485 & $-$27.97588 & fsrq & 0.926 & PKS 2255$-$282 & 1.3 \\
	\hline
	\hline
	J0102.8+5825 & 15.71134 & 58.41576 & fsrq & 0.644 & TXS 0059+581 & 2.1 \\
	J0252.8$-$2218 & 43.20377 & $-$22.32386 & fsrq & 1.419 & PKS 0250$-$225 & 1.2 \\
	J0303.4$-$2407 & 45.86259 & $-$24.12074 & bll & 0.266 & PKS 0301$-$243 & 2 \\
	J0428.6$-$3756 & 67.17261 & $-$37.94081 & bll & 1.11 & PKS 0426$-$380 & 3.4 \\ 
	J0449.4$-$4350 & 72.36042 & $-$43.83719 & bll & 0.205 & PKS 0447$-$439 & 2.5 \\ 
	J0457.0$-$2324 & 74.26096 & $-$23.41384 & fsrq & 1.003 & PKS 0454$-$234 & 2.6 \\
	J0501.2$-$0157 & 75.30886 & $-$1.98359 & fsrq & 2.291 & S3 0458$-$02 & 1.7\\
	J0721.9+7120 & 110.48882 & 71.34127 & bll & 0.127 & S5 0716+71 & 2.8 \\
	J0818.2+4223 &	124.56174 & 42.38367 & bll &  0.530	& S4 0814+42 & 2.2 \\
	J1303.0+2435 & 195.75454 & 24.56873 & bll & 0.993 & MG2 J130304+2434 & 2 \\
	J1649.4+5238 & 252.35208 & 52.58336 & bll & -- & 87GB 164812.2+524023 & 2.7 \\
	J1903.2+5541 & 285.80851 & 55.67557 & bll & -- & TXS 1902+556 & 3.8 \\
	J2056.2$-$4714 & 314.06768 & $-$47.23386 & fsrq & 1.489 & PKS 2052$-$47 & 1.7 \\
	\enddata
\end{deluxetable*}

\begin{deluxetable*}{lccccccc|c|cc}[h]
	\tablecaption{List of periods and their associated confidence level$\slash$FAP for the 11 periodic-emission candidates and the 13 low significance candidates in Table~\ref{tab:candidates_list}. There are some sources with two periods with high significance (organized by amplitude of the peak). Additionally, the average of the periods from different methods and their significance levels are shown. The \myhash ~ denotes that the LC of the object is an evenly-spaced LC, therefore the wavelet period is generated by the CWT method. The symbol $\dagger$ denotes the PDM results that present the effects described in $\S$\ref{sec:sub_candidates_b}. Finally, stars denote sources whose LCs were previously studied in a similar context by other authors, the column ``Literature'' shows the period previously reported. Note that the REDFIT method only gives a maximum of significance of $2.5\sigma$ (see text for details). All periods are in years. \label{tab:a_candiadtes_periods}}
	\tabletypesize{\footnotesize} 
	\tablewidth{0pt}
	\tablehead{
		\colhead{Association Name} &
		\colhead{Power-Law} &
		\colhead{Boostrap} & 
		\colhead{Simulated LC} &
		\colhead{REDFIT} &
		\colhead{PDM} &
		\colhead{WWZ} &	
		\colhead{DFT-Welch} &
		\colhead{Average} &	
		\colhead{Literature} &
		\\
		\colhead{} &
		\colhead{} &
		\colhead{} &
		\colhead{} &
		\colhead{} &
		\colhead{} &
		\colhead{\myhash CWT} &
		\colhead{} &
		\colhead{} &
		\colhead{} &
	}
	\startdata		   
	\hline
	GB6 J0043+3426  & 1.9 ($>$4$\sigma$) & 1.9 ($>$4$\sigma$) & 1.8 ($>$4$\sigma$) & 2 ($>$1.5$\sigma$) & $\dagger$1.8 ($\approx$4$\sigma$) & 1.9 ($>$4$\sigma$) & 1.7 ($\approx$4$\sigma$) & 1.8 ($\approx$4$\sigma$) & - \\
	PKS 0208$-$512 & \makecell{2.6 ($\approx$3$\sigma$) \\ 0.9 ($\approx$3$\sigma$) \\ 1.3 ($\approx$3$\sigma$)} & 2.7 ($\approx$4$\sigma$) & \makecell{2.6 ($>$4$\sigma$) \\ 0.9 ($\approx$4$\sigma$) \\ 1.3 ($\approx$4$\sigma$)}& \makecell{2.6 ($\approx$2.5$\sigma$) \\ 0.9 ($>$2$\sigma$)} & 2.6 ($\approx$4$\sigma$) & 2.7 ($\approx$4$\sigma$) & 2.1 ($>$2$\sigma$) & 2.6 ($>$3$\sigma$) & - \\	
	MG1 J021114+1051 & \makecell{1.8 ($\approx$3$\sigma$) \\ 0.8 ($\approx$3$\sigma$)} & 1.7 ($\approx$4$\sigma$) & \makecell{1.8 ($>$2$\sigma$) \\ 0.8 ($>$2$\sigma$)} & 1.8 ($>$2.5$\sigma$) & 1.7 ($\approx$4$\sigma$) & 1.8 ($>$4$\sigma$) & 1.5 ($\approx$3) & 1.7 ($>$3.5$\sigma$) & - \\
	TXS 0518+211 & 2.9 ($\approx$2$\sigma$) & 2.9 ($>$4$\sigma$) & 2.9 ($\approx$2$\sigma$) & 2.9 ($>$2$\sigma$) & 2.5 ($\approx$4$\sigma$) & 3 ($>$4$\sigma$) & 2.6 ($\approx$4$\sigma$) & 2.8 ($>$3$\sigma$)  & - \\	
	OJ 014 & 4.2 ($\approx$1$\sigma$) & 4.1 ($>$4$\sigma$) & 4.6 ($>$4$\sigma$) & 4.6 ($>$2.5$\sigma$) & 4.4 ($\approx$4$\sigma$)  & 4.4 ($\approx$4$\sigma$)  & 3.7 ($\approx$4$\sigma$) & 4.3 ($>$3.5$\sigma$) & - \\ 	
	S4 1144+40 & 3.4 ($\approx$2$\sigma$) & 3.5 ($>$4$\sigma$) & 3.4 ($>$2$\sigma$) & 3.4 ($>$2.5$\sigma$) & 3.3 ($\approx$4$\sigma$) & 3.5 ($\approx$4$\sigma$) & 3.2 ($\approx$4$\sigma$) & 3.3 ($>$3$\sigma$) & - \\ 
	PG 1246+586 & 2 ($>$2$\sigma$) & 2 ($\approx$4$\sigma$) & 2 ($>$1$\sigma$) & 2.2 ($\approx$2.5$\sigma$) & $\dagger$3.9 ($\approx$4$\sigma$) & 2 ($\approx$4$\sigma$) & 2.3 ($\approx$4$\sigma$) & 2 ($\approx$3$\sigma$) & - \\	
	TXS 1452+516 & 2 ($>$3$\sigma$) & 2 ($>$4$\sigma$) & \makecell{2.3 ($>$4$\sigma$) \\ 1 ($>$4$\sigma$) \\ 1.6 ($>$4$\sigma$)} & \makecell{2.3 ($>$2$\sigma$) \\ 1 ($\approx$2$\sigma$) \\ 1.6 ($>$2$\sigma$)} & $\dagger$4.9 ($\approx$4$\sigma$) & 2.2 ($>$4$\sigma$) & 1.8 ($\approx$4$\sigma$) & 2.1 ($>$3.5$\sigma$) & - \\
	PG 1553+113* & 2.2 ($>$4$\sigma$) & 2.2 ($>$4$\sigma$) & 2.2 ($>$4$\sigma$) & 2.1 ($>$2.5$\sigma$) & 2.2 ($\approx$4$\sigma$) & \myhash2.3 ($>$3$\sigma$) & 2.3 ($\approx$4$\sigma$) & 2.2 ($>$4$\sigma$) & 2.2 \\	
	PKS 2155$-$304* & 1.7 ($\approx$3$\sigma$) & 1.7 ($>$4$\sigma$) & 1.5 ($\approx$2$\sigma$) & 1.7 ($>$2.5$\sigma$) & $\dagger$1.7 ($\approx$4$\sigma$) & \myhash1.7 ($\approx$2$\sigma$) & 1.5 ($\approx$4$\sigma$) & 1.7 ($>$3$\sigma$) & 1.7 \\
	PKS 2255$-$282 & 1.3 ($\approx$3$\sigma$) & 1.3 ($>$4$\sigma$) & 1.3 ($\approx$4$\sigma$) & 1.3 ($>$2.5$\sigma$) & $\dagger$2.7 ($\approx$4$\sigma$) & 1.4 ($>$4$\sigma$) & 1.4 ($\approx$4$\sigma$) & 1.3 ($>$3.5$\sigma$) & - \\
	\hline
	\hline
	TXS 0059+581 & 2.2 ($\approx$2$\sigma$) & 2.2 ($>$3$\sigma$) & 2.2 ($>$3$\sigma$) & 2.2 ($>$1.5$\sigma$) & $\dagger$4.2 ($\approx$4$\sigma$) & 2.2 ($\approx$4$\sigma$) & 1.8 ($\approx$2$\sigma$) & 2.1 ($\approx$3$\sigma$) & - \\	
	PKS 0250$-$225  & 1.2 ($\approx$4$\sigma$) & 1.2 ($\approx$2$\sigma$) & 1.2 ($\approx$2$\sigma$) & 1.2 ($>$2.5$\sigma$) & 1.2 ($\approx$3$\sigma$) & 1.2 ($>$4$\sigma$) & 1.4 ($\approx$2$\sigma$) & 1.2 ($>$2.5$\sigma$) & - \\
	PKS 0301$-$243* & 2 ($\approx$3$\sigma$) & 2.1 ($>$1$\sigma$) & 2 ($>$2$\sigma$) & 2 ($\approx$2.5$\sigma$) & 4.5 ($\approx$4$\sigma$)  & 2.1 ($>$4$\sigma$)  & 2 ($\approx$3$\sigma$) & 2 ($\approx$3$\sigma$) & 2.1 \\
	PKS 0426$-$380* & \makecell{3.4 ($>$1$\sigma$) \\ 1 ($>$2$\sigma$)} & \makecell{3.9 ($>$4$\sigma$) \\ 1 ($\approx$4$\sigma$) \\ 1.3 ($\approx$3$\sigma$) \\ 1.6($\approx$3$\sigma$)} & 3.4 ($>$4$\sigma$) & 3.4 ($>$2$\sigma$) & 3.5 ($\approx$4$\sigma$) & \myhash3 ($>$2$\sigma$) & 1.3 ($\approx$2$\sigma$) & 3.4 ($\approx$3$\sigma$) & 3.3 \\	
	PKS 0447$-$439 & \makecell{2.5 ($\approx$3$\sigma$) \\ 1.2 ($>$2$\sigma$)} & \makecell{2.5 ($>$4$\sigma$) \\ 1.7 ($>$2$\sigma$)} & \makecell{2.5 ($\approx$3$\sigma$) \\ 1.2 ($\approx$4$\sigma$) \\ 1.7 ($>$2$\sigma$)} & \makecell{2.5 ($\approx$2$\sigma$) \\ 1.2 ($\approx$2$\sigma$)} & 2.5 ($\approx$4$\sigma$) & \myhash2.2 ($\approx$2$\sigma$) & 2.4 ($\approx$4$\sigma$) & 2.5 ($\approx$3$\sigma$) & - \\
	PKS 0454$-$234  & 2.4 ($\approx$3$\sigma$) & 2.4 ($\approx$4$\sigma$) & 2.4 ($>$2$\sigma$) & 3.5 ($\approx$1.5$\sigma$) & 2.5 ($\approx$4$\sigma$) & 2.4 ($\approx$4$\sigma$) & 2.4 ($>$2$\sigma$) & 2.6 ($>$2.5$\sigma$) & - \\	
	S3 0458$-$02  & 1.7 ($\approx$4$\sigma$) & 1.7 ($\approx$2$\sigma$) & 1.7 ($\approx$2$\sigma$) & 1.7 ($>$2.5$\sigma$) & 1.9 ($\approx$2$\sigma$) & 1.8 ($>$4$\sigma$) & 1.6 ($\approx$3$\sigma$) & 1.7 ($>$2.5$\sigma$) & - \\
	S5 0716+71* & \makecell{ 2.7 ($\approx$3$\sigma$) \\ 0.9 ($\approx$3$\sigma$)} & 2.7 ($>$2$\sigma$) & \makecell{2.9 ($>$2$\sigma$)\\0.9 ($>$4$\sigma$)} & \makecell{2.7 ($\approx$2.5$\sigma$) \\ 0.9 ($\approx$2$\sigma$)} & \makecell{2.7 ($\approx$4$\sigma$) \\ 0.9 ($\approx$2$\sigma$)} & \makecell{\myhash3 ($>$2$\sigma$) \\ \myhash0.9 ($\approx$2$\sigma$)} & 3 ($\approx$4$\sigma$) & 2.8 ($>$2.5$\sigma$) & 0.9 \\	
	S4 0814+42 & 2.2 ($\approx$3$\sigma$) & 2.1 ($>$2$\sigma$) & 2.1 ($>$4$\sigma$) & 2.2 ($\approx$2.5$\sigma$) & $\dagger$ 4.1 ($\approx$4$\sigma$) & 2.2 ($\approx$4$\sigma$) & 4.5 ($\approx$2$\sigma$) & 2.2 ($\approx$3.5$\sigma$) & - \\
	MG2 J130304+2434 & 2 ($>$2$\sigma$) & 1.9 ($>$3$\sigma$) & 2 ($\approx$3$\sigma$) & 1.9 ($>$2$\sigma$) & $\dagger$4 ($\approx$4$\sigma$) & 2 ($\approx$4$\sigma$) & 2.2 ($\approx$2$\sigma$) & 2 ($>$2.5$\sigma$) & - \\
	87GB 164812.2+524023 & 2.7 ($\approx$3$\sigma$) & 2.7 ($>$3$\sigma$) & 2.7 ($>$2$\sigma$) & 2.7 ($>$2.5$\sigma$) & 2.7 ($>$2$\sigma$) & 2.7 ($\approx$4$\sigma$) & 1.6 ($>$2$\sigma$) & 2.7 ($>$2.5$\sigma$) & - \\
	TXS 1902+556 & 3.8 ($\approx$2$\sigma$) & 3.7 ($\approx$3$\sigma$) & 3.8 ($>$1$\sigma$) & 3.8 ($>$2.5$\sigma$) & 3.7 ($\approx$4$\sigma$) & 3.8 ($\approx$4$\sigma$) & 3.5 ($\approx$2$\sigma$) & 3.8 ($>$2.5$\sigma$) & - \\
	PKS 2052$-$47* & \makecell{1.8 ($>$2$\sigma$) \\ 2.8 ($\approx$2$\sigma$)}& \makecell{1.7 ($>$3$\sigma$) \\ 2.8 ($>$3$\sigma$)} & \makecell{1.8 ($\approx$4$\sigma$) \\ 2.8 ($\approx$4$\sigma$)}  & \makecell{1.8 ($>$2$\sigma$) \\ 2.8 ($>$2$\sigma$)} & 1.7 ($\approx$4$\sigma$) & 2.7 ($\approx$4$\sigma$) & 1.5 ($\approx$2$\sigma$) & 1.7 ($>$2.5$\sigma$) & 1.7 \\
	\enddata
\end{deluxetable*}

\begin{deluxetable*}{lccccccc}[h]
	\tablecaption{List of periods provided by the MCMC and Bayesian-QPO methods for the 11 periodic-emission candidates and the 13 low significance candidates in Table~\ref{tab:candidates_list}. Additionally, the sensitivity of the QPO method (see $\S$\ref{sec:bayesian}) and B-LTR values are included. An {\it X} represents that the algorithm did not converge and thus no value was reported in the frequency range considered ($\S$\ref{sec:bayesian}). The column ``Flares'' denotes sources that clearly have high activity (flaring states, according to the methodology described in $\S$\ref{sec:others}). All periods are in years. \label{tab:a_candidates_ts}} 
	\tablewidth{0pt}
	\tablehead{
		\colhead{Association Name} &
		\colhead{MCMC Sine Fitting} &
		\colhead{Bayesian} &
		\colhead{Maximum Sensitivity} &
		\colhead{B-LRT} &
		\colhead{Flares} &
	}
	\startdata
	GB6 J0043+3426 & $2.1^{+0.1}_{-0.4}$  & $\approx$1.8 & $\approx$26\% & 0.01\% & - \\
	PKS 0208$-$512 & 2.7 $\pm$ 0.1 & $\approx$2.7 & $\approx$40\% & 3.7\% & \redcheck \\
	MG1 J021114+1051 & 1.8 $\pm$ 0.1 & $\approx$1.5 & $\approx$31.6\% & 46.8 \% & \redcheck \\
	TXS 0518+211 & 2.9 $\pm$ 0.1 & $\approx$3 & $\approx$52.4\% & 8.5\% & - \\	
	OJ 014 & 4.6 $\pm$ 0.2 & $\approx$0.28 & \%55 & 6.5\% & - \\
	S4 1144+40 & 3.5 $\pm$ 0.1 & $\approx$3.8 & $\approx$113\% & 78.9\% & - \\
	PG 1246+586 & $2.2^{+2.6}_{-0.1}$ & $\approx$2.3 & $\approx$20\% & 0.3\% & -\\
	TXS 1452+516 & 2.2 $\pm$ 0.1 & $\approx$1.9 & $\approx$46\% & 0.01\% & - \\
	PG 1553+113 & 2.2 $\pm$ 0.1 & $\approx$2.3 & $\approx$35\% & 3.5\% & - \\
	PKS 2155$-$304 & 1.7 $\pm$ 0.1 & $\approx$1.8 & $\approx$18\% & 0.01\% & - \\		
	PKS 2255$-$282 & 1.3 $\pm$ 0.1 & $\approx$3.8 & $\approx$52\% & 0.01\% & - \\
	\hline
	\hline
	TXS 0059+581 & $2.2^{+1.5}_{-0.1}$ & 2.4 & $\approx$41\% & 99.5\% & \redcheck \\	
	PKS 0250$-$225 & 1.2 $\pm$ 0.1 & X & X & 0.01\% & \redcheck \\
	PKS 0301$-$243 & 2.1 $\pm$ 0.1 & $\approx$2.1 & $\approx$31\% & 0.5\% & \redcheck \\
	PKS 0426$-$380 & \makecell{3.2 $\pm$ 0.1 \\ 1.7 $\pm$ 0.1} & $\approx$3.8 & $\approx$70\% & 6.5\% & - \\
	PKS 0447$-$439 & \makecell{2.5 $\pm$ 0.1 \\ 1.2 $\pm$ 0.1} & $\approx$2.4 & $\approx$27.3\% & 0.2\% & - \\
	PKS 0454$-$234 & 2.3 $\pm$ 0.1 & $\approx$3.8 & $\approx$51.6\% & 10.4\% & - \\
	S3 0458$-$02 & 1.8 $\pm$ 0.1 & $\approx$3.8 & $\approx$26\% & 18.3\% & \redcheck \\
	S5 0716+71 & $2.7^{+0.1}_{-1.8}$ & $\approx$3.8 & $\approx$71\% & 6.1\% & - \\
	S4 0814+42 & $2.8^{+1.1}_{-0.1}$ & $\approx$2.8 & $\approx$23\% & 3.4\% & - \\
	MG2 J130304+2434 & $2.1^{+0.1}_{-0.7}$ & $\approx$2.3 & $\approx$60\% & 0.01\% & \redcheck \\
	87GB 164812.2+524023 & 2.8 $\pm$ 0.1 & $\approx$1.6 & $\approx$67.7\% & 0.01\% & - \\
	TXS 1902+556 & 3.7 $\pm$ 0.2 & $\approx$3.4 & $\approx$19.0\% & 6.2\% & - \\
	PKS 2052$-$47 & \makecell{$1.7^{+0.1}_{-0.4}$ \\ $2.6^{+0.1}_{-0.8}$} & X & X & 0.01\% & \redcheck \\
	\enddata
\end{deluxetable*}
\begin{deluxetable*}{lccccc}[h]
	\tablecaption{Estimates of the observational requirements to reach 5$\sigma$ in the period detection for the 24 sources in Table~\ref{tab:candidates_list}. The estimation is represented by the average of number of cycles and the additional years of the LAT observations associated to such cycles. The LCs employed in this study are 9 year-long, from August 2008 until October 2017. The starting moment for the extra years of the LAT observations is considered October 2017. The \textit{X} values denotes that it was not possible to obtain the estimation. \label{tab:estimation_5sigma}}
	\tablewidth{0pt}
	\tablehead{
		\colhead{Association Name} &
		\colhead{\myhash of Cycles to Reach 5$\sigma$} &
		\colhead{\myhash Extra Years of LAT Observation } &
		\colhead{\myhash Total Years of LAT Observation } &
	}
	\startdata
	GB6 J0043+3426 & $\approx$2.8 & $\approx$6 & $\approx$15 \\
	TXS 0059+581 & X & X & X \\	
	PKS 0208$-$512 & $\approx$1.6 & $\approx$4 & $\approx$13 \\
	MG1 J021114+1051 & $\approx$2 & $\approx$4 & $\approx$13 \\
	PKS 0250$-$225 & $\approx$3 & $\approx$3 & $\approx$12 \\
	PKS 0301$-$243 & $\approx$3.6 & $\approx$7 & $\approx$16 \\
	PKS 0426$-$380 & $\approx$2 & $\approx$6 & $\approx$15 \\
	PKS 0447$-$439 & $\approx$2 & $\approx$5 & $\approx$14 \\
	PKS 0454$-$234 & $\approx$1.4 & $\approx$4 & $\approx$13 \\
	S3 0458-02 & X & X & X \\
	TXS 0518+211 & $\approx$1.4 & $\approx$4 & $\approx$13 \\	
	S5 0716+71 & X & X & X \\
	S4 0814+42 & $\approx$2 & $\approx$4.4 & $\approx$13 \\
	OJ 014 & $\approx$2.9 & $\approx$12 & $\approx$21 \\	
	S4 1144+40 & $\approx$1.9 & $\approx$ 7 & $\approx$15 \\
	PG 1246+586 & $\approx$2.2 & $\approx$4 & $\approx$13 \\
	MG2 J130304+2434 & X & X & X \\
	TXS 1452+516 & $\approx$1.4 & $\approx$3 & $\approx$12 \\
	PG 1553+113 & $\approx$ 1 & $\approx$2 & $\approx$ 11 \\
	87GB 164812.2+524023 & $\approx$ 1.4 & $\approx$4 & $\approx$13 \\
	TXS 1902+556 & X & X & X \\
	PKS 2052$-$47 & $\approx$ 2.8 & $\approx$5 & $\approx$14 \\
	PKS 2155$-$304 & $\approx$ 2 & $\approx$3.4 & $\approx$13 \\
	PKS 2255$-$282 & $\approx$ 1.8 & $\approx$ 2.5 & $\approx$11.5 \\		\enddata
\end{deluxetable*}

\end{document}